\documentclass[a4paper,11pt]{article}
\usepackage{jinstpub} 
\usepackage{lineno}

\title{Broad-band, high-gain, low-frequency Antennas for Radio Detection of Earth-skimming Tau Neutrinos}

\author[a,b]{T. Huege\note{Corresponding author.}}
\author[c]{and O. Krömer}
\affiliation[a]{Institute for Astroparticle Physics (IAP), Karlsruhe Institute of Technology, Karlsruhe, Germany}
\affiliation[b]{Astrophysical Institute, Vrije Universiteit Brussel, Brussels, Belgium}
\affiliation[c]{Institute for Data Processing and Electronics (IPE), Karlsruhe Institute of Technology, Karlsruhe, Germany}

\emailAdd{tim.huege@kit.edu}

\abstract{A promising approach to detect high-energy tau neutrinos is through the measurement of impulsive radio emission from horizontal air showers initiated in the Earth's atmosphere. Observations at frequencies between 30 and 80\,MHz seem particularly promising -- if high-gain antennas focused at the horizon and blocking out as much as possible of the noisy sky are employed. Due to the large wavelengths, however, designing an antenna with the required properties is highly non-trivial at such low frequencies. In this article, we explore suitable antenna designs that provide the desired high gain, possess a smooth beam, are insensitive to ground conditions, are easily impedance-matched over the wide band, and are mechanically simple for deployment in large numbers in inaccessible terrain. In particular, we consider the ``rhombus'' antenna design for both horizontally and vertically polarized radiation a very attractive option for tau neutrino detection efforts in the atmosphere with the radio technique.}


\arxivnumber{1234.56789} 

\begin{document}
\maketitle
\flushbottom


\section{Introduction}
\label{sec:intro}

With the neutrino flux at PeV energies firmly established \cite{IceCube:2013low,IceCube:2024fxo}, the race for the detection of astrophysical neutrinos at even higher energies is on. One promising approach employs the radio detection of Earth-skimming tau neutrinos that interact in the Earth's crust or mountains, with the resulting tau lepton exiting the medium and then decaying in the air, thereby initiating an extensive air shower. Through the geomagnetic and Askaryan mechanisms, the resulting air shower will emit pulsed radio signals at frequencies of tens to hundreds of MHz \cite{Huege:2016veh}, detectable with radio-antenna arrays. The air showers of interest will be coming from the horizon, will be slightly upward-going, and will be best-detectable at distances of tens of kilometres, where a good compromise is achieved between detection threshold (the signal is diluted with distance) and size of the illuminated area (the emission is forward-beamed in a cone of only a few degrees opening angle). Prominent contenders of this approach are the GRAND \cite{GRAND:2018iaj} and BEACON \cite{Southall:2022yil} projects followed up recently with ideas aiming to combine the two approaches \cite{Wissel:2024rhp}.

Of the possible frequency bands for the detection of radio signals from such air showers the one below 100\,MHz, typically between 30 and 80\,MHz is of particular interest. At these frequencies the forward-beamed coherent emission smoothly illuminates a relatively broad cone \cite{Wissel:2020sec}, which is favorable for the detection with a distributed antenna array, whereas at higher frequencies the signal is much more focused on a narrow Cherenkov ring, reducing the probability that a particular antenna in a given array is illuminated. Also, readout electronics for frequencies below 100\,MHz are less expensive than at higher frequencies, resulting data rates are lower, and time-synchronization to within one nanosecond, i.e., within 30$^\circ$ of phase as needed for interferometric beam-forming of distributed detectors, has been successfully demonstrated already using beacon reference transmitters \cite{Schroder:2010sa,PierreAuger:2015aqe,PierreAuger:2023opi}.

Of course, the low frequencies also bring significant disadvantages. The first one is that, with a 10\,m wavelength at 30\,MHz, antennas for this frequency band become large, posing a challenge for the mechanical design and associated costs. The second one is that the Galactic background at these frequencies has a temperature of order 10,000\,K and is thus strongly dominating over system noise \cite{Huege:2016veh,SchroederReview}. The third disadvantage is that at the low edge of this frequency band, radio-frequency interference from far-away sources can be reflected off the ionosphere, especially during high solar activity \cite{DiegoSolar2024}.

In this article, we present possible antenna designs tailored at the radio detection of extensive air showers from Earth-skimming tau neutrinos. In section \ref{sec:criteria} we discuss the criteria for a suitable antenna design, before discussing particular candidate designs in the following sections.




\section{Criteria for suitable antenna designs} \label{sec:criteria}

An antenna lending itself well to the detection of extensive air showers arising from Earth-skimming tau neutrinos should fulfill the following criteria:

{\bf High gain:} The air showers of interest will arise from a small and defined solid angle, from the horizon or from mountains at the horizon. To maximize the signal-to-noise ratio, it is desirable that the antenna has a very high gain and is sensitive only to that region of interest, and not the remainder of the sky. In particular, as the sky is very ``loud'' at these low frequencies, blocking as much of it as possible will be very beneficial. (The noise temperature of the ground is an open question; depending on its specific characteristics it might have a significant reflectivity and reflection of the sky might result in noise temperatures much higher than the nominal 300\,K ground temperature.) The ideal antenna will thus in particular have a small field-of-view in elevation (maybe 5-10$^\circ$), while the desired field-of-view in the azimuth direction depends on the experimental design. It could either be desired to have it relatively wide (lower gain, larger instantaneous field of view) or narrow (higher gain, smaller instantaneous field of view).

{\bf Large bandwidth:} The canonical band of 30\,to 80\,MHz is favorable because at lower frequencies atmospheric noise (reflection of signals from far-away sources on the ionosphere) becomes too high, whereas above 80\,MHz often the FM radio band precludes effective measurements. The antenna should cover this whole band with a relatively uniform response; sensitivity up to even higher frequencies can be an advantage for remote sites where the FM band is not present.

{\bf Smooth gain pattern:} It is desirable that the gain pattern of the antenna is smooth and featureless with as little structure and frequency dependence as possible within the region of interest, and with only few and well-suppressed sidelobes outside of it.

{\bf Impedance matching:} Ideally, the antenna is easy to match with a canonical 50\,$\Omega$ readout chain over its full (broad) band, thereby allowing optimum exploitation of the signal, not suffering from losses due to internal reflections.

{\bf Stability of response} The antenna should, ideally, have a relatively stable response that does not change strongly with parameters such as ground conductivity (wet versus dry ground), slight bending of wires, slight misorientation, slight variation of antenna height above ground, etc.

{\bf Simple mechanical design:} An antenna can fulfill all of the above criteria but be completely impractical for cost or mechanical reasons. A practical antenna should use a minimum amount of material, should not require a very complex mounting (several concrete foundations, very high masts, ...) and should in particular not exhibit a too-high wind load. It also needs to be durable in harsh and inaccessible environments on timescales of 10+ years.

{\bf Low dispersion:} To maximize the signal-to-noise ratio for a radio-based self-trigger, the antenna should exhibit low dispersion, i.e., not spread the pulse in time more than necessary. That said, dispersion can be compensated for ``online'' before making a trigger decision \cite{Schmidt:2011zzc}. While dispersion would complicate the trigger design, low dispersion is thus not as important as the other criteria listed above, and we thus did not investigate it specifically in this work.

With these critera in mind, we have explored various options for suitable antenna designs, which we will discuss now in the following. All of the simulations were carried out with the antenna modeling program EZNEC+ v.5.0. If not stated otherwise, we assumed a ground with a conductivity of $\sigma = 0.005$\,S/m and a dielectricity constant of $\epsilon_\mathrm{r}=13$, which corresponds to regular, slightly humid ground.


\section{Rhombus antennas}

A very promising design that lends itself well to both the detection of horizontal-polarisation (H-pol) and vertical-polarisation (V-pol) signals is that of an aperiodic rhombus antenna. It consists of a ``rhombus'' made of wires suspended at a height of a few metres above the ground, terminated at the end to which the main lobe is pointing, opposite the readout terminals.

\subsection{Horizontal rhombus antenna} \label{sec:nominalrhombus}

\begin{figure}
\centering
\includegraphics[width=0.52\textwidth]{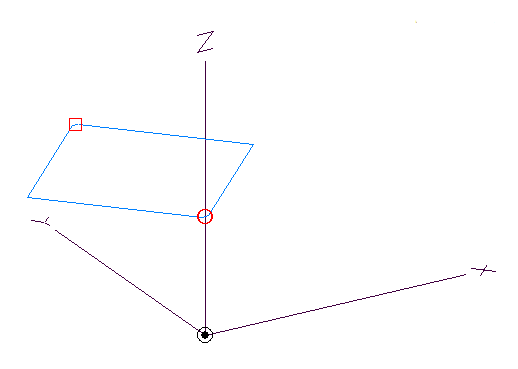}
\includegraphics[width=0.47\textwidth]{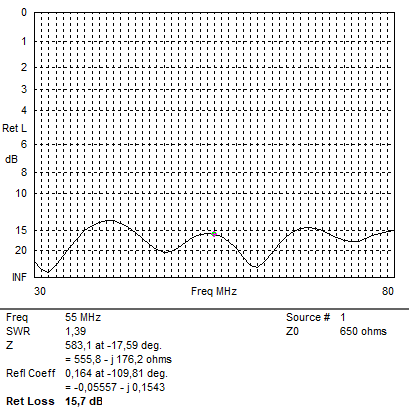}
\includegraphics[width=0.48\textwidth]{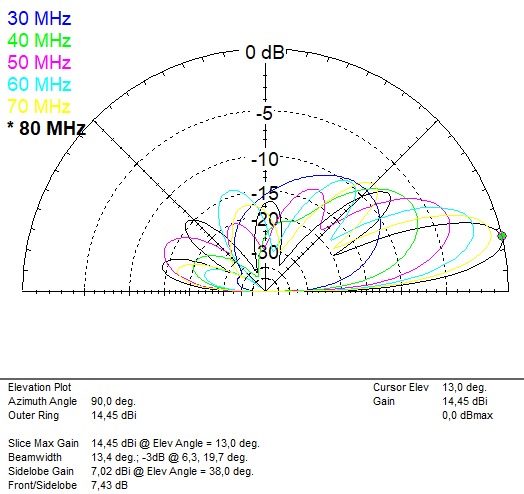}
\includegraphics[width=0.49\textwidth]{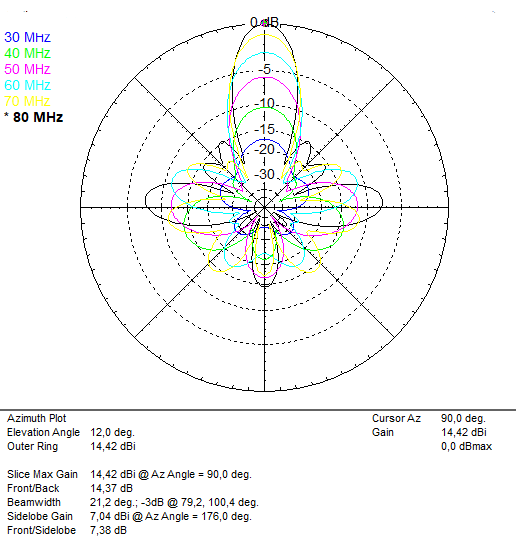}
\caption{Simulation of a horizontally polarized aperiodic rhombus antenna, see text for details. Top left: drawing of the antenna design, top right: S11 return loss, bottom left: gain pattern in elevation, bottom right: gain pattern in azimuth. The bottom panels illustrate the gain patterns for various frequencies, the color code is listed in the figure. The details listed below the diagrams refer to the frequency of 80 MHz.} \label{fig:rhombus-hpol}
\end{figure}

The design of a rhombus antenna for H-pol signals is depicted in the top-left panel of figure \ref{fig:rhombus-hpol}. The antenna consists of four wires with a length of 5.6\,m each, making 90$^\circ$ angles with each other, thus forming a square. The wires have a thickness of 4\,mm and are suspended 4\,m above ground. Ideally, one would prepare one foundation and then suspend all four corners of the rhombus with non-conducting (e.g., fiberglass) rods from this one foundation. The readout terminal is at the position marked with the circle. At the position marked with the square, the antenna is terminated with a 750\,$\Omega$ resistor. The main lobe of the antenna faces into the y-direction.

The S11 return loss of the antenna shown in the top-right panel of figure \ref{fig:rhombus-hpol} provides information on the ease of impedance matching over the bandwidth of 30\,to 80\,MHz to a real impedance such as a 50\,$\Omega$ readout chain or the optimum generator impedance $R_{G,opt.}$ of the first RF transistor if noise matching is used. The lower and flatter the return loss, the easier a good wide-band matching with minimum reflections is achieved. The return loss of the H-pol rhombus antenna is less than 15 dB over the whole band. The radiation resistance $Z_0$ corresponds to a real impedance of about 600\,$\Omega$ with a small inductive or capacitive fraction. Consequently, the antenna can be easily matched to the readout chain with a balun transformer which also converts the differential antenna terminal to a single-ended signal. This unpretentious wideband matching strategy may also be applied to all other antennas presented in the following. 

The bottom-left panel of figure \ref{fig:rhombus-hpol} shows the antenna gain as a function of frequency from 30 to 80\,MHz with further quantitative details listed for the 80\,MHz case. Especially at the higher end of the band, the main lobe of the antenna is narrow in elevation with a half-power beam width of 13.4$^\circ$, pointing at 13$^\circ$ above the ground. On a ground sloped by 15-20$^\circ$, e.g., on a mountain slope, the main lobe will thus point towards the horizon as desired. The gain at 80\,MHz reaches 14.4\,dBi with the strongest sidelobe suppressed by more than 7\,dB. Also the backward direction is strongly suppressed. At lower frequencies, the main lobe becomes broader and points further up, especially at the lower end of 30\,MHz. This behavior is not desired but could potentially be improved by slight design changes, see below.

The bottom-right panel of figure \ref{fig:rhombus-hpol} illustrates the azimuthal gain pattern at frequencies from 30 to 80\,MHz for an elevation of 12.0$^\circ$, within the main lobe. The azimuthal beam is very narrow with a width of only 21$^\circ$, which will lead to a relatively small instantaneous field of view yet with a very high gain.

\subsection{Variations of the horizontal rhombus antenna}

The wires for the horizontal rhombus antenna need not be oriented at a 90$^\circ$ angle. We also simulated a design consisting of four wires with a length of 5.0\,m each making 108$^\circ$ angles at the sides. This results in an antenna with a footprint of 6\,m by 8\,m. Overall, the performance is similar with a slightly decreased main-lobe gain of 12.8\,dBi.

Variation of the height of the wires above ground influences the width of the main lobe and thus affects the maximum gain. For wires suspended 2\,m above ground the beam width in elevation at 80\,MHz increases to 23$^\circ$ and the gain drops to 12.7\,dBi. For a rhombus at 6\,m above ground the beam narrows to 9$^\circ$ with a maximum gain of 14.8\,dBi. At the same time, the side lobes increase with the wire-height above ground, at 6\,m the strongest sidelobe is suppressed by only 2.5\,dB.

The height of the wires above ground also determines the elevation at which the beam has its maximum gain. For 2\,m this is at 25$^\circ$ while for 6\,m this drops to 10$^\circ$. For a given detector site with a particular slope the height above ground can thus be used to tune the beam geometry as needed (within limits). Reasonable wire heights above ground are between 3\,m and 5\,m.

\begin{figure}
\centering
\includegraphics[width=0.5\textwidth]{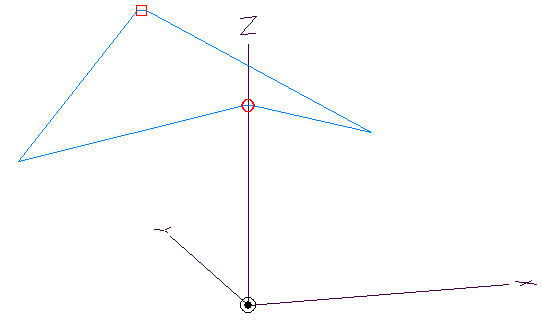}
\includegraphics[width=0.42\textwidth]{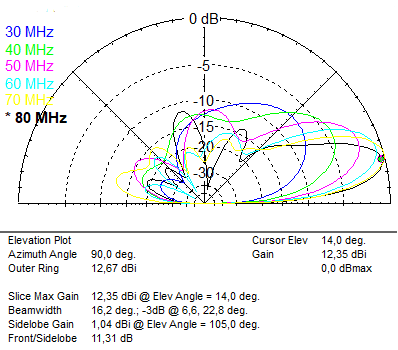}
\caption{Left: drawing of a horizontally-polarized aperiodic rhombus antenna with improved low-frequency behaviour, Right: corresponding elevation gain pattern with details for the frequency of 80\,MHz.} \label{fig:rhombus-hpol-lowfreq}
\end{figure}

If desired, the low-frequency response can be optimized by placing the termination and readout terminal at 4.5\,m above ground and lowering the other two edges of the rhombus to 2.5\,m as shown in figure \ref{fig:rhombus-hpol-lowfreq}. This way, at the lower end of the band the gain is increased and the elevation of the main lobe points further down. Furthermore, side lobe suppression is increased.

The termination resistance influences the forward-to-backward gain ratio, the value of 750\,$\Omega$ has the optimum suppression in backward direction. In principle, a complex impedance instead of an Ohmic resistor can be used to tune the frequency dependence of the antenna, as could be a quarter wave stub.

\begin{figure}
\centering
\includegraphics[width=0.52\textwidth]{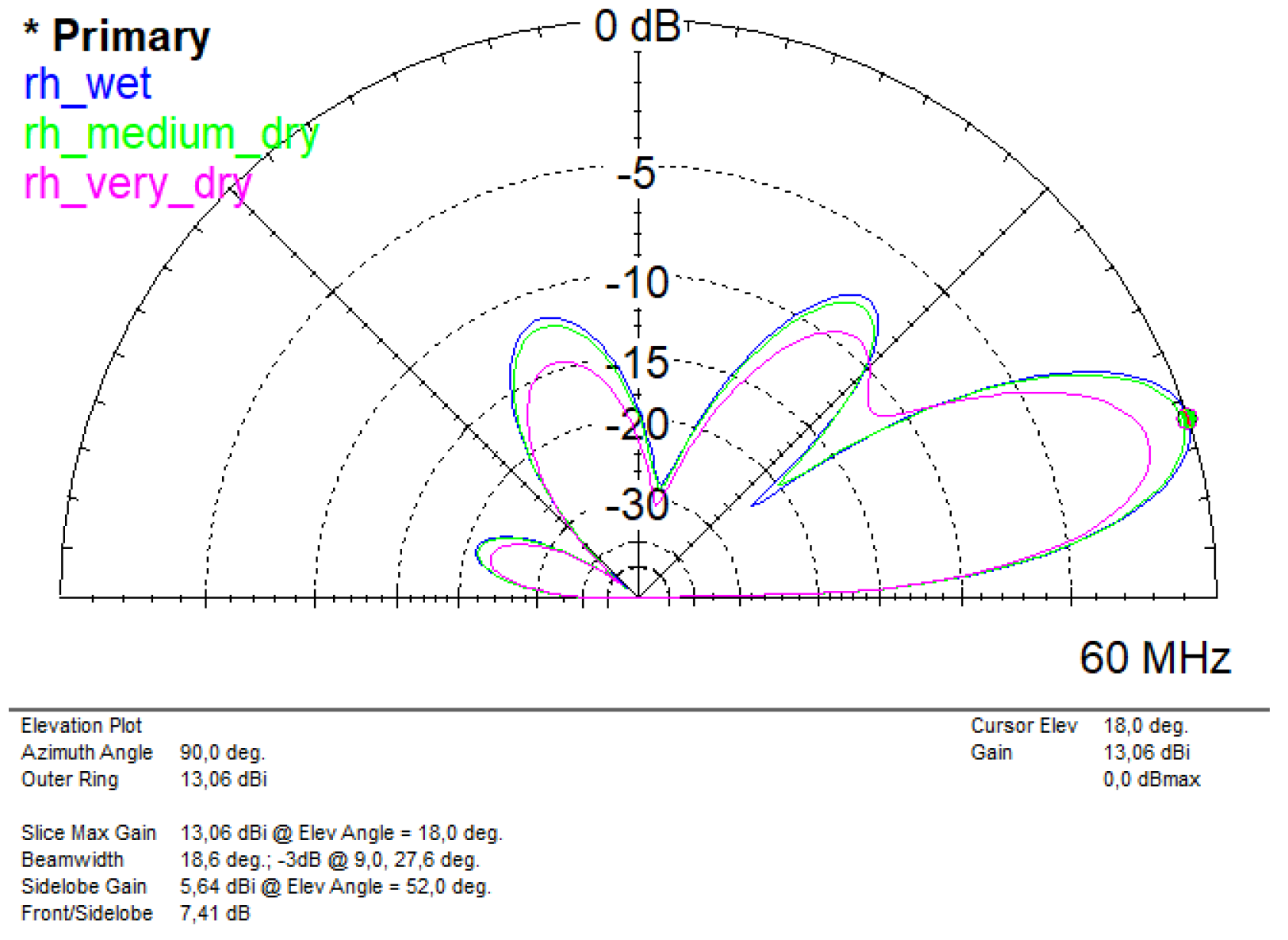}
\caption{Simulation of the ground dependence of the gain pattern in elevation for a horizontally polarized aperiodic rhombus antenna.} \label{fig:rhombus-hpol-ground}
\end{figure}

We also investigated the influence of different ground conditions, from very dry ($\sigma = 0.0001$\,S/m and $\epsilon_\mathrm{r}=3$) to wet ($\sigma = 0.01$\,S/m and $\epsilon_\mathrm{r}=30$) on the nominal antenna design described in section \ref{sec:nominalrhombus}. The gain pattern at 60\,MHz, shown in figure \ref{fig:rhombus-hpol-ground}, remains very stable for all ground conditions, except for a loss of about 1.5\,dBi maximum gain for very dry ground.

As the radiation resistance of the antenna is relatively high with a value of 600\,$\Omega$, the choice of material (e.g., aluminum versus copper) as well as the diameter of the wires (we tested variations from 2\,mm to 10\,mm) has no relevant impact. Furthermore, we validated with simulations that four vertical masts suspending the wires do not have a relevant impact on the antenna response even if they are made of conducting material. Finally, we varied the inclination of the horizontal rhombus against the ground with angles varying from -10$^\circ$ to +10$^\circ$. Perhaps somewhat unintuitively, the influence of this is negligible, which means that minor imprecisions during deployment do not have a negative impact.

\subsection{Vertical rhombus antenna}

\begin{figure}
\centering
\includegraphics[width=0.52\textwidth]{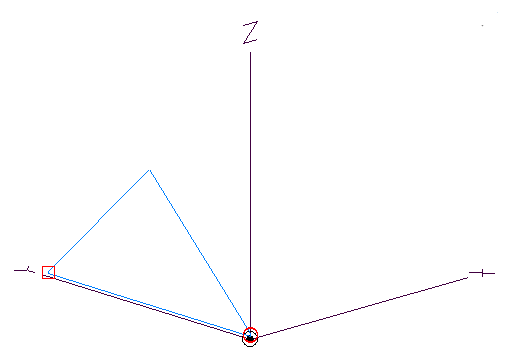}
\includegraphics[width=0.47\textwidth]{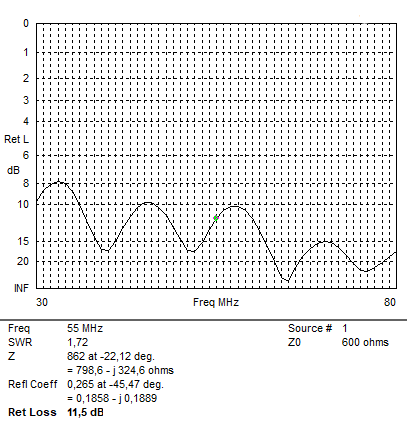}
\includegraphics[width=0.48\textwidth]{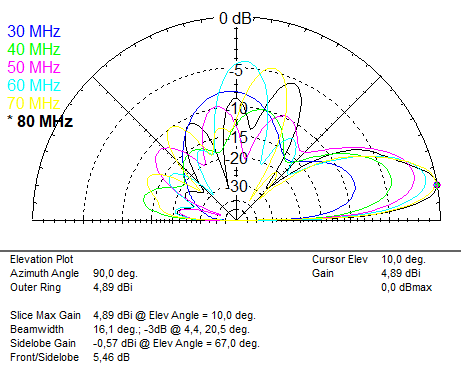}
\includegraphics[width=0.49\textwidth]{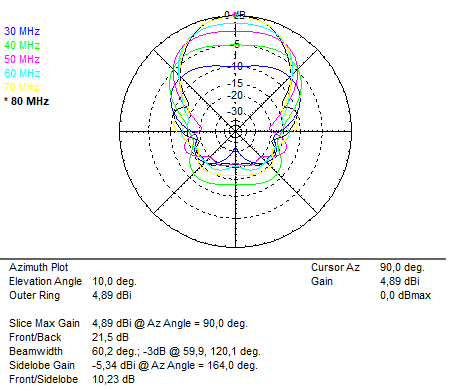}
\caption{Simulation of a vertically polarized aperiodic rhombus antenna, see text for details. Top left: drawing of the antenna design, top right: S11 return loss, bottom left: gain pattern in elevation, bottom right: gain pattern in azimuth. The bottom panels illustrate the gain patterns for various frequencies, the color code is listed in the figure. The details listed below the diagrams refer to the frequency of 80 MHz.} \label{fig:rhombus-vpol}
\end{figure}

The vertically polarized rhombus antenna is depicted in the top-left panel of figure \ref{fig:rhombus-vpol}. It consists of only three wires, as the ground effectively ``mirrors'' the triangular structure. The wire running on the ground has a length of 10\,m, the other two wires have a length of 7.07\,m each and make an angle of 90\,$^\circ$ at the top corner, which is suspended 5\,m above the ground with a single non-conducting mast. Again, the read-out terminal is at the position marked with the circle, whereas a 350\,$\Omega$ resistor is terminating the wires at the position marked with the square. The S11 return loss shown in the top-right panel illustrates that reflections are less than 10 dB over the whole bandwidth at a real impedance of 600\,$\Omega$. Again, a balun transformer may be used for relatively simple wideband impedance transformation to the readout chain.

The elevation dependence of the antenna gain, shown in the bottom-left panel, shows that at 80\,MHz, the main lobe points at 10.0$^\circ$ above the ground, appropriate for a sloped surface, with a width of 16$^\circ$. The first sidelobe is suppressed by more than 5\,dB. Again, for the low frequencies the main beam is wider and in this case exhibits a prominent sidelobe towards the zenith. The azimuthal gain pattern, shown in the bottom-right panel at an elevation of 15$^\circ$ has a width of 75$^\circ$ and thus is much broader than for the H-pol rhombus antenna. Consequently, the maximum achievable gain is significantly lower than for the H-pol rhombus, at 4.9\,dBi instead of 14.4\,dBi. Nevertheless, the easy mechanical design and simple matching make it an attractive antenna, especially if a wider instantaneous field-of-view is desired.

\section{Logarithmic-Periodic Dipole antennas}

Logarithmic Periodic Dipole Antennas (LPDAs) have been used in the field of radio detection of extensive air showers for a long time \cite{PierreAuger:2012ker}, albeit with their main lobe oriented towards the zenith. Except for their dispersive nature (the phase center for each dipole and thus resonant frequency is at a different location introducing a frequency-dependent group delay), they generally have favorable characteristics, i.e., they have a relatively flat response over a broad frequency band, a high gain, allow simple impedance matching, are relatively independent of ground conditions, and possess a mostly featureless gain pattern. Here, we explore a horizontally and a vertically polarized LPDA as a candidate antenna for the detection of extensive air showers from Earth-skimming tau neutrinos.

\subsection{Horizontal LPDA}

\begin{figure}
\centering
\includegraphics[width=0.52\textwidth]{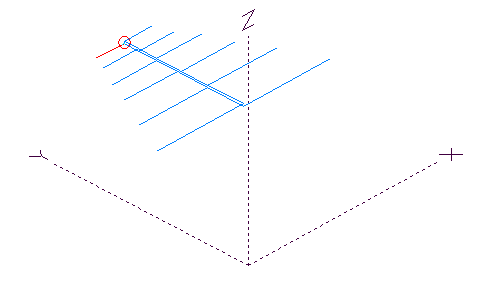}
\includegraphics[width=0.47\textwidth]{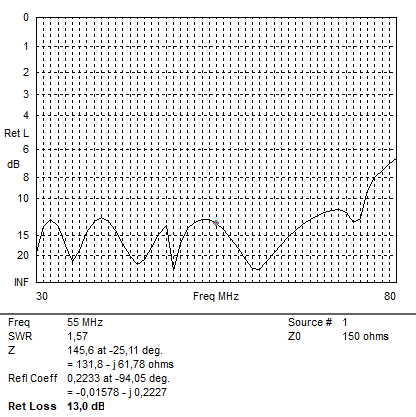}
\includegraphics[width=0.48\textwidth]{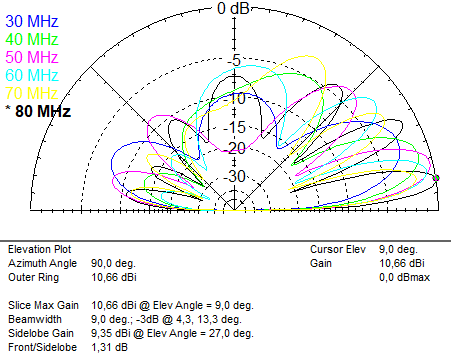}
\includegraphics[width=0.49\textwidth]{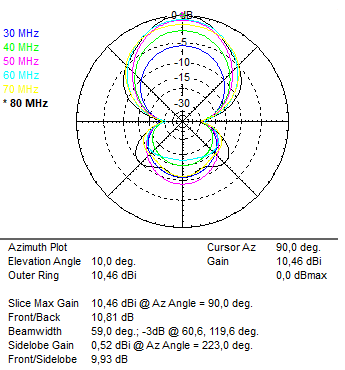}
\caption{Simulation of a horizontally polarized LPDA, see text for details. Top left: drawing of the antenna design, top right: S11 return loss, bottom left: gain pattern in elevation, bottom right: gain pattern in azimuth. The bottom panels illustrate the gain patterns for various frequencies, the color code is listed in the figure. The details listed below the diagrams refer to the frequency of 80 MHz.} \label{fig:lpda-hpol}
\end{figure}

In figure \ref{fig:lpda-hpol} we show the characteristics of a horizontally polarized LPDA consisting of 6 individual dipoles, positioned 6 meters above ground, with the readout terminals at the front of the antenna. The longest dipole has a length of 5\,m and the length of the boom is 3.5\,m. The maximum gain amounts to 10.7\,dBi at an elevation of 9$^\circ$, with a beam width of also 9$^\circ$. While the gain pattern of an LPDA directed at the zenith is featureless, ground reflections introduce significant structure in the elevation-dependence of the gain for a horizontally polarized LPDA. The most prominent sidelobe at 80\,MHz is suppressed by a mere 1.3\,dB with respect to the main lobe. The width of the main lobe in azimuth is of order 60$^\circ$ at 80\,MHz. Impedance matching to a 50\,$\Omega$ readout chain is unproblematic.

In addition to the undesired sidelobe structure in the elevation dependence of the gain pattern, the sheer size of the LPDA is non-ideal from a mechanical point of view. Potentially, there are ways to miniaturize LPDA antenna designs without compromising performance too much \cite{StephDielectric,s21186034,8016657}, which would be worthwhile to investigate further.

\subsection{Vertical LPDA}

\begin{figure}
\centering
\includegraphics[width=0.52\textwidth]{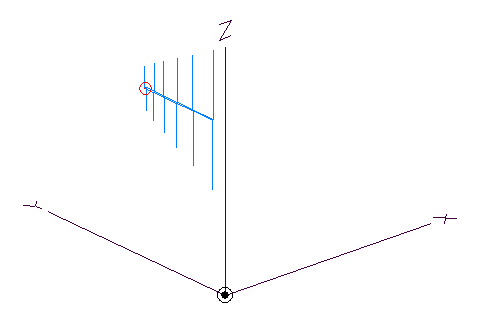}
\includegraphics[width=0.47\textwidth]{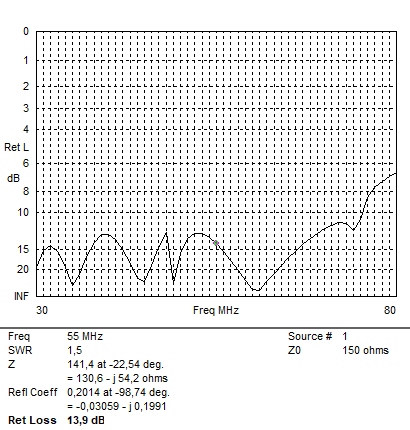}
\includegraphics[width=0.48\textwidth]{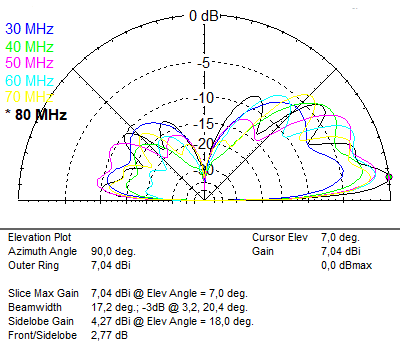}
\includegraphics[width=0.49\textwidth]{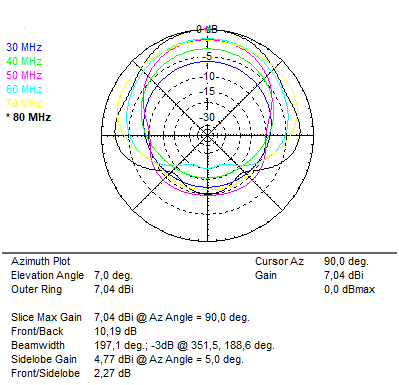}
\caption{Simulation of a vertically polarized LPDA, see text for details. Top left: drawing of the antenna design, Top right: S11 return loss, bottom left: gain pattern in elevation, bottom right: gain pattern in azimuth. The bottom panels illustrate the gain patterns for various frequencies, the color code is listed in the figure. The details listed below the diagrams refer to the frequency of 80 MHz.} \label{fig:lpda-vpol}
\end{figure}

A vertically aligned LPDA, again consisting of 6 dipoles distributed over a boom length of 3.5\,m, with the longest dipole having a length of 5\,m, and the horizontal structure mounted at 6\,m above ground, has more favorable characteristics than the horizontally-polarized version, as shown in figure \ref {fig:lpda-vpol}. The maximum gain at 80\,MHz is 7.0\,dBi at an elevation of 7$^\circ$ and with a width in elevation of 17$^\circ$. Towards lower frequencies, the gain reduces and the beam gets wider. The most prominent sidelobe at 80\,MHz is suppressed by about 2.8\,dB. The azimuthal gain pattern shows a beam width of order $180^\circ$.

Again, the sheer size of the LPDA is unfavorable, especially since no (conducting) structures must overlap with the antenna structure. A metal mast should be offset from the lowest-frequency dipole by 2\,m, adding to the mechanical stress imposed by the design. Again, miniaturisation could be an option to explore.

\section{Stacked Short Aperiodic Loaded Loop antennas}

Another antenna design that has proven favorable for the detection of air-shower radio signals is that of the Short Aperiodic Loaded Loop Antenna (SALLA) \cite{KroemerSALLA2009}. This design has been employed successfully both in the Tunka-Rex experiment \cite{TUNKA-REX:2013fao} as well as the Auger Radio Detector consisting of more than 1,600 SALLAs on an area of 3,000\,km$^2$ \cite{Pont:2021pwd}. In these experiments, where the SALLA is pointed at the zenith, the antenna has a very homogeneous and broad gain pattern over the band from 30-80\,MHz, it is easily impedance-matched, mechanically simple, and the resistor at the bottom makes it hardly sensitive to ground conditions or structures below the antenna. The disadvantage of the SALLA is that, due to the resistor, significant noise is introduced, which is, however, not so much of a concern when the very ``loud'' sky in the 30-80\,MHz band dominates the noise budget. For an antenna trying to block out the sky and profit from the lower effective temperature of the ground, this is, however, of course very unattractive. Nevertheless, the SALLA antenna could be a useful element in a design phasing together several antennas. In the following, we present a concept for a design consisting of three ``stacked'' SALLA antennas as a basis for potential further developments. For the simulations shown here the stacked SALLAs are read out in phase with balanced amplitude. Combining the SALLAs with a defined phase shift and amplitude ratio might be an option for further optimization.

\subsection{Horizontal stacked SALLAs}

\begin{figure}
\centering
\includegraphics[width=0.52\textwidth]{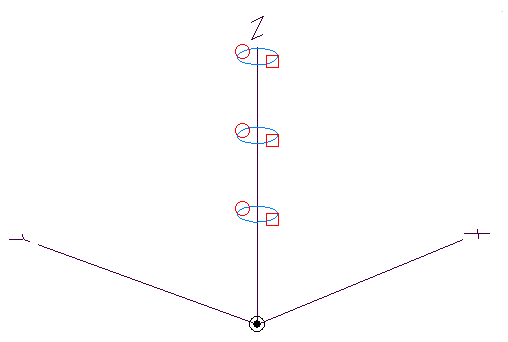}
\includegraphics[width=0.47\textwidth]{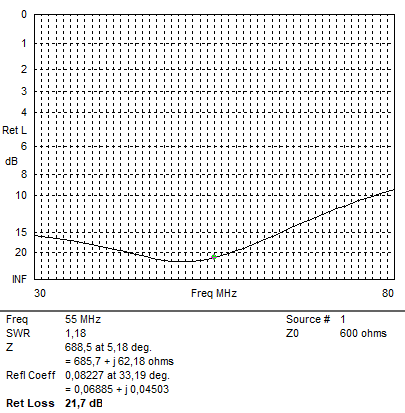}
\includegraphics[width=0.48\textwidth]{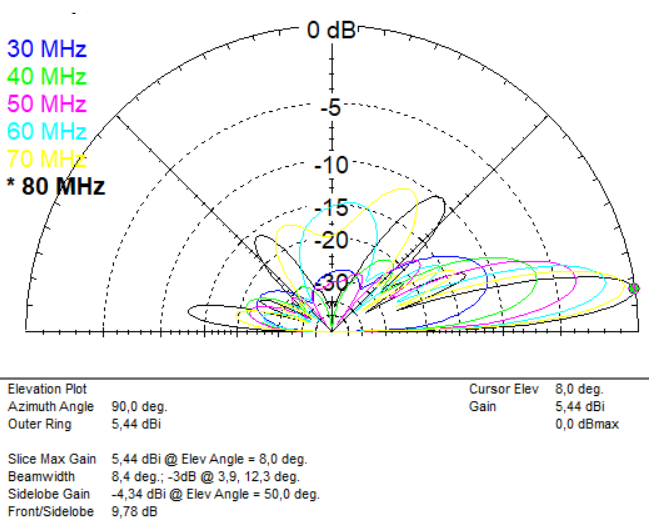}
\includegraphics[width=0.49\textwidth]{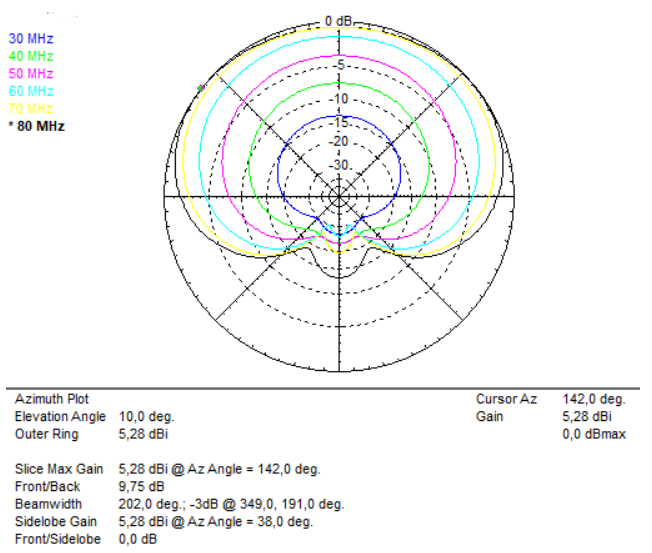}
\caption{Simulation of horizontally polarized stacked SALLAs, see text for details. Top left: drawing of the antenna design, top right: S11 return loss, bottom left: gain pattern in elevation, bottom right: gain pattern in azimuth. The bottom panels illustrate the gain patterns for various frequencies, the color code is listed in the figure. The details listed below the diagrams refer to the frequency of 80 MHz.} \label{fig:salla-hpol}
\end{figure}

For the case of horizontally polarized radiation, we investigated a design of three horizontally oriented SALLAs with a diameter of 1.2\,m each positioned at 3.5\,m, 6.0\,m and 8.5\,m above ground, cf.\ figure \ref{fig:salla-hpol}. The three resistors in the backward direction amount to 800\,$\Omega$. While this design requires a high mast, it is otherwise relatively simple from a mechanical point of view. The gain pattern is favorable. The main lobe at a frequency of 80\,MHz exhibits a maximum gain of 5\,dBi at an elevation of 8$^\circ$ and possesses a width of 8$^\circ$. The most prominent side-lobe is suppressed by approximately 10\,dB. In the azimuthal direction the main beam is larger than 180$^\circ$, the backward direction is very strongly suppressed. Matching to a 50\,$\Omega$ readout chain is unproblematic.

\subsection{Vertical stacked SALLAs}

\begin{figure}
\centering
\includegraphics[width=0.52\textwidth]{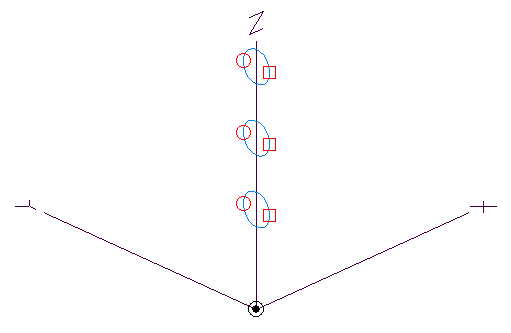}
\includegraphics[width=0.47\textwidth]{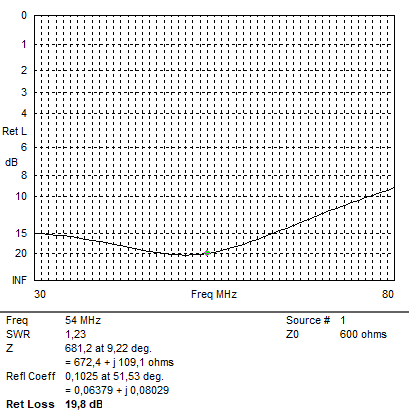}
\includegraphics[width=0.48\textwidth]{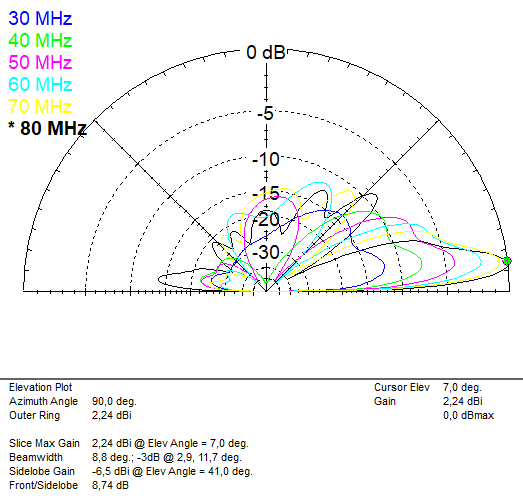}
\includegraphics[width=0.49\textwidth]{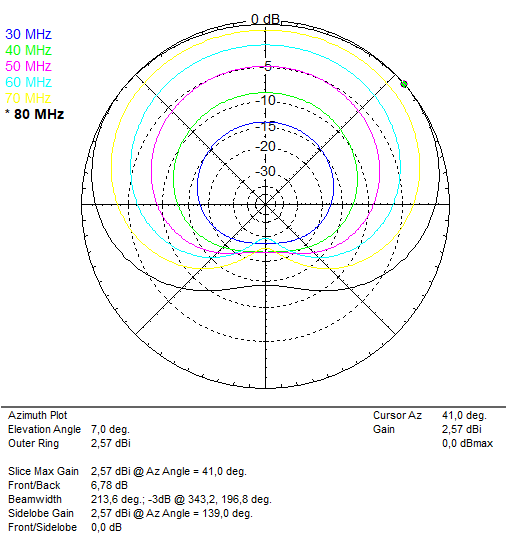}
\caption{Simulation of vertically polarized stacked SALLAs, see text for details. Top left: drawing of the antenna design, top right: S11 return loss, bottom left: gain pattern in elevation, bottom right: gain pattern in azimuth. The bottom panels illustrate the gain patterns for various frequencies, the color code is listed in the figure. The details listed below the diagrams refer to the frequency of 80 MHz.} \label{fig:salla-vpol}
\end{figure}

In figure \ref{fig:salla-vpol} we show the response for three vertically-oriented stacked SALLAs, again with a diameter of 1.2\,m, a resistor of 800\,$\Omega$, and mounted at heights of 3.5\,m, 6.0\,m and 8.5\,m above ground. The gain pattern is again favorable, with a main beam at 80\,MHz at an elevation of around 7$^\circ$ and a beamwidth of around 7.5$^\circ$ in elevation. However, the maximum gain is only around 2.5\,dBi. The azimuthal gain pattern is smooth and larger than 180$^\circ$. Impedance matching is unproblematic.

\section{Other explored designs}

We have explored a number of other possible designs, including vertical ``fan dipoles'', straight or tilted, and an aperiodically terminated long wire sensitive to vertically polarized emission. The ``fan dipoles'' reach maximum gains of 9 and 11\,dBi, but are difficult to impedance-match and also have non-favorable gain patterns. The aperiodic long wire can reach a maximum gain of 7\,dBi and can be easily impedance-matched, but shows an unfavorable amount of side lobes.

\section{Conclusions}

We have explored a number of antenna designs for mechanically simple high-gain antennas for the 30-80\,MHz band. Such antennas are an important component for experiments aiming at the radio detection of air showers arising from Earth-skimming tau neutrinos. Out of the explored designs we consider the ``rhombus antenna'' particularly attractive. It is mechanically very simple and has a narrow beam at a low elevation above the horizon so that, deployed on a mountain slope, the beam would point to the desired small band of elevation angles near the horizon. The horizontally polarized version of the rhombus antenna has a narrow beam in the azimuthal direction of around 20$^\circ$ half-power width and consequently a very high gain of more than 14\,dBi at 80\,MHz. The vertically polarized version has a wider beam in azimuth and otherwise similarly attractive characteristics. Both are easily impedance-matched to a 50\,$\Omega$ readout chain and are relatively insensitive to ground conditions and slight changes of the geometry.

Logarithmic-periodic dipole antennas are also an option, although at these frequencies they become large and thus pose mechanical issues. The horizontally polarized version also suffers from significant sidelobes due to ground reflections, while the vertically polarized version is more benign. If they could be miniaturized (e.g., through the use of a dielectric), they might yet be an attractive design. Another option we explored is a vertical stack of three short aperiodic loaded loop antennas, which exhibits favorable gain patterns but a relatively low overall gain, and requires a very high mast.

With the rhombus antennas in particular, we have provided an attractive antenna design for radio detection of Earth-skimming tau neutrinos through tau-induced extensive air showers, which can be further fine-tuned for application in a given experimental setup.

\acknowledgments

The authors would like to thank Kumiko Kotera, Olivier Martineau and Stephanie Wissel for helpful discussions.











\providecommand{\href}[2]{#2}\begingroup\raggedright\endgroup

\end{document}